# A Supervised Machine Learning Framework for Multipactor Breakdown Prediction in High-Power Radio Frequency Devices and Accelerator Components: A Case Study in Planar Geometry

Asif Iqbal[a*], John Verboncoeur[b#], and Peng Zhang[a+]

[a]Department of Nuclear Engineering and Radiological Sciences, University of Michigan, Ann Arbor, MI 48109, USA

[b]Department of Computational Mathematics, Science, and Engineering, Michigan State University, East Lansing, MI 48824, USA

*asifiq@umich.edu, #johnv@msu.edu, +umpeng@umich.edu

Corresponding author:Asif Iqbal(asifiq@umich.edu),Peng Zhang(umpeng@umich.edu)

## Abstract

Multipactor is a nonlinear electron avalanche phenomenon that can severely impair the performance of high-power radio frequency (RF) devices and accelerator systems. Accurate prediction of multipactor susceptibility across different materials and operational regimes remains a critical yet computationally intensive challenge in accelerator component design and RF engineering. This study presents the first application of supervised machine learning (ML) for predicting multipactor susceptibility in two-surface planar geometries. A simulation-derived dataset spanning six distinct secondary electron yield (SEY) material profiles is used to train regression models—including Random Forest (RF), Extra Trees (ET), Extreme Gradient Boosting (XGBoost), and funnel-structured Multilayer Perceptrons (MLPs)—to predict the time-averaged electron growth rate, $\delta_{\text{avg}}$. Performance is evaluated using Intersection over Union (IoU), Structural Similarity Index (SSIM), and Pearson correlation coefficient. Tree-based models consistently outperform MLPs in generalizing across disjoint material domains. MLPs trained using a scalarized objective function that combines IoU and SSIM during Bayesian hyperparameter optimization with 5-fold cross-validation outperform those trained with single-objective loss functions. Principal Component Analysis reveals that performance degradation for certain materials stems from disjoint feature-space distributions, underscoring the need for broader dataset coverage. This study demonstrates both the promise and limitations of ML-based multipactor prediction and lays the groundwork for accelerated, data-driven modeling in advanced RF and accelerator system design.

# 1. Introduction

Multipactor discharge (Iqbal et al., 2018; Kishek et al., 1998; Vaughan, 1988) is a vacuum electron avalanche that occurs in high-power radio frequency (RF) and microwave (HPM) devices (Benford et al., 2025). This highly nonlinear phenomenon is triggered by secondary electron emission (SEE) (Furman and Pivi, 2002; Iqbal et al., 2019; Ludwick et al., 2020; Vaughan, 1989, 1993) and can develop under a range of conditions, including resonant interactions between electrons and the RF electric field in two-surface geometries (e.g., planar (Iqbal et al., 2022; Vaughan, 1988), stripline (Mirmozafari et al., 2022, 2021)), and coaxial (Langellotti et al., 2021; Woo, 1968a, 1968b) vacuum RF components), as well as non-resonant avalanche modes such as single-surface multipactor (Iqbal et al., 2018; Kishek and Lau, 1998; Zhang et al., 2011), which typically occurs in components like dielectric windows.

Multipactor discharge is a persistent challenge in high-power RF systems and accelerator components, where it contributes to power loss, localized heating, surface degradation, vacuum instability, and component failure (González-Iglesias et al., 2024; Iqbal et al., 2023a; Jing et al., 2013; Vaughan, 1988). In the Spallation Neutron Source (SNS) at Oak Ridge National Laboratory, multipactor-induced electron activity in the vacuum environment has been linked to RF window failures and ceramic disk heating (Macek et al., 2003; Toby et al., 2023, 2021). At the Large Hadron Collider (LHC), multipacting contributes to pressure rise, beam instabilities, and cryogenic heat loads (Bruning et al., 1999), while in the Linac Coherent Light Source II-High Energy (LCLS-II-HE) cryomodules at the Stanford Linear Accelerator Center (SLAC), multipactor has been associated with cavity quench events at high gradients (Giaccone et al., 2022; Posen et al., 2022). Similar concerns have been reported in Fermilab's Proton Improvement Plan-II (PIP-II) (Romanov, 2017) and at the Los Alamos Neutron Science Center (LANSCE) linear accelerators, where multipactor prompted RF injector redesigns (Simakov,Evgenya et al., 2024; Xu,Haoran et al., 2024). These examples underscore the urgent need for predictive multipactor modeling to improve reliability and guide mitigation strategies in next-generation high power RF and accelerator systems.

Prior to the advent of computational tools and techniques, multipactor prediction approaches largely relied on experimental investigations (Farnsworth, 1938; Preist and Talcott, 1961; Vaughan, 1961) and analytical models (Gill and von Engel, 1948; Gutton and Gutton, 1928; Hatch and Williams, 1958, 1954; Vaughan, 1988) that described basic mechanisms of secondary electron avalanches. However, these models often made simplifying assumptions about geometry, field structure, and emission dynamics, limiting their applicability to practical devices with complex shapes and variable operating conditions. With advances in computational capabilities, more sophisticated techniques such as Monte Carlo (MC) (Iqbal et al., 2020; Kishek, 1997), statistical (Anza et al., 2012, 2010; Lin et al., 2023; Vdovicheva et al., 2004), and Particle-in-Cell (PIC) simulations (Vahedi et al., 1993; Vahedi and DiPeso, 1997; Verboncoeur et al., 1993; Xiao et al., 2019) were developed to capture the stochastic and dynamic nature of multipactor growth. PIC simulations, in particular, have become powerful tools for modeling time-dependent electron motion and space-charge effects in realistic electromagnetic fields. Nevertheless, MC, PIC, and statistical methods are computationally intensive, especially when applied to large-scale 3D geometries or when conducting extensive parametric sweeps over frequency, voltage, material properties, and device configurations. Even with parallel computing and high-performance simulation frameworks, multipactor prediction in practical devices remains a time-consuming and resource-demanding task (Iqbal et al., 2023a), making rapid design optimization or real-time susceptibility evaluation challenging for many applications.

Machine learning (ML) techniques offer powerful tools for modeling complex, nonlinear physical systems where traditional analytical methods or simulations become impractical. Recent studies have demonstrated the effectiveness of ML models in accelerating predictions in fields such as accelerator technology (Edelen et al., 2018; Kaiser et al., 2024; Pathak, 2024), plasma physics (Dalsania et al., 2021; "Special issue," 2023; Spirkin et al., 2020; Zhong et al., 2020), fluid dynamics (Tizakast et al., 2023), and electromagnetics (Jeong et al., 2024; Li et al., 2021; Sagar et al., 2021; Zou et al., 2024).

This work, to our knowledge, presents the first application of machine learning for predicting multipactor susceptibility. By training surrogate supervised ML models to learn the mapping between input features (e.g., applied voltage, frequency-gap distance product, and secondary electron emission properties of materials; to be discussed in detail in Section 2) and susceptibility outcomes in a two-surface planar geometry, we develop a data-driven framework that enables rapid susceptibility prediction without the need for repeated, computationally expensive simulations, offering significant reductions in computational cost while maintaining high fidelity in predictive tasks. Beyond demonstrating the feasibility of ML for this application, this study lays the groundwork for future benchmarking efforts and accelerated model development in multipactor research.

The remainder of this paper is organized as follows. Section 2 details the dataset generation procedure based on high-fidelity 3D Particle-in-Cell (PIC) simulations, defines the input features and target variable, and outlines the proposed machine learning workflow, including the evaluation metrics and feature relevance analysis, introduces the machine learning models investigated—linear regression, tree-based ensemble methods, and neural networks—and presents the model selection rationale and cross-validation framework. Section 4 evaluates the predictive performance of the models, analyzes susceptibility region matching across materials, and discusses generalization challenges.

Section 5 concludes the study by summarizing the key findings and offering perspectives for future research on machine learning-driven multipactor prediction.

## 2. Materials and methods

*2.1 Multipactor Susceptibility Dataset via 3D PIC Simulations*

Although prior simulation and experimental studies on planar multipactor discharge exist, standardized datasets suitable for machine learning applications remain scarce. To address this gap, we constructed a high-fidelity, simulation-based dataset tailored for supervised learning. The data were generated using three-dimensional (3D) electromagnetic Particle-in-Cell (PIC) simulations (Iqbal et al., 2023b, 2022) performed in CST Particle Studio (CST PS) ("CST Studio Suite," 2019). The simulation domain consisted of a parallel-plate geometry, excited by a 1 GHz time-harmonic RF signal applied across a variable vacuum gap distance ($d$). Secondary electron yield (SEY) was modeled using Vaughan's SEY model (Vaughan, 1993) as implemented in CST PS, with emission energy and angular distributions of secondary electrons following the approach described by Iqbal et al. (2023b, 2022). Space-charge effects were turned off, allowing the investigation of pure secondary electron-driven multipactor growth without space-charge suppression (Iqbal et al., 2023b, 2022). For each combination of RF voltage amplitude $V_{rf}$ and gap distance $d$, the multipactor susceptibility parameter, defined as the time-averaged electron growth rate $\delta_{avg}$, was computed as described by Iqbal et al. (2022) over a simulation duration of 10 ns (approximately 10 RF cycles).

Multipactor susceptibility is typically presented as a two-dimensional mapping of $\delta_{avg}$ in the ($V_{rf}$, $f \times d$) plane (Fig. 1), following the similarity scaling law of Vaughan (1988), where $f$ is the excitation frequency. Regions where $\delta_{avg} > 1$ represent sustained multipactor growth and susceptibility to breakdown, whereas regions where $\delta_{avg} < 1$ correspond to stable, multipactor-free conditions. These susceptibility charts (Fig. 1) form the basis for training the machine learning models to predict multipactor onset under varying material and operational parameters.

The dataset encompasses six different material property sets, each characterized by distinct secondary electron yield (SEY) parameters, namely, the maximum SEY ($\delta_{max0}$) at normal incidence of electrons onto a surface, the incident energy at maximum SEY ($E_{max0}$), and the first and second crossover energies ($E_1$ and $E_2$, respectively), where SEY equals unity. These material parameters were chosen according to Vaughan's model (Vaughan, 1989) to represent a range of hypothetical surfaces encountered in high-power RF and accelerator components (Table 1).

Table 1: Secondary electron yield (SEY) parameters for the six material property sets used in this study.

| Material | $\delta_{max0}$ | $E_{max0}$ (eV) | $E_1$ (eV) | $E_2$ (eV) |
|---|---|---|---|---|
| M1 | 2.09 | 165.0 | 18.0 | 1900.0 |
| M2 | 2.09 | 277.5 | 42.0 | 3056.0 |
| M3 | 2.09 | 400.0 | 44.0 | 4604.0 |
| M4 | 1.2 | 277.5 | 109.5 | 759.5 |
| M5 | 1.2 | 400.0 | 158.0 | 1094.0 |
| M6 | 3.2 | 400.0 | 19.0 | 1550.0 |

The applied RF voltage, $V_{rf}$, as well as the gap distance, $d$ (and consequently, the frequency-gap distance product, $fd$), were systematically varied across a broad range to cover the relevant operating regimes for practical applications (Fig. 1). Simulation outputs were compiled into a single database, with each entry corresponding to a unique combination of material properties, RF voltage, and frequency-gap distance product. This dataset was then used for feature extraction and model training.

For each material property set, susceptibility data were directly simulated on a $51 \times 24$ grid over the ($V_{rf}$, $fd$) plane, spanning $V_{rf} \in [30,18000]$ V and $fd \in [0.7,20]$ GHz·mm, yielding 1224 simulation points per chart. Each Particle-in-Cell (PIC) simulation point required approximately 3-5 minutes of compute time on an average, resulting in a total simulation duration of about 72-100 hours per chart. To augment the dataset (Kearney et al., 2023; Khamlich et al., 2025; Shorten and Khoshgoftaar, 2019; Yang et al., 2024; Zhu et al., 2023) while minimizing the risk of introducing non-physical artifacts, nonlinear spline interpolation was applied strictly within the bounds of the simulated ($V_{rf}$, $fd$) domain (i.e., no extrapolation), yielding 10,000 points per chart. This strategy enabled the generation of a denser dataset, substantially reducing the computational burden while maintaining high fidelity to the original PIC results.

The simulations were conducted on a workstation equipped with two 12-core Intel Xeon E5-2687W v4 processors. Despite the use of powerful multi-core computing hardware, the simulation campaign spanned several weeks, reflecting the significant computational cost associated with 3D PIC modeling.

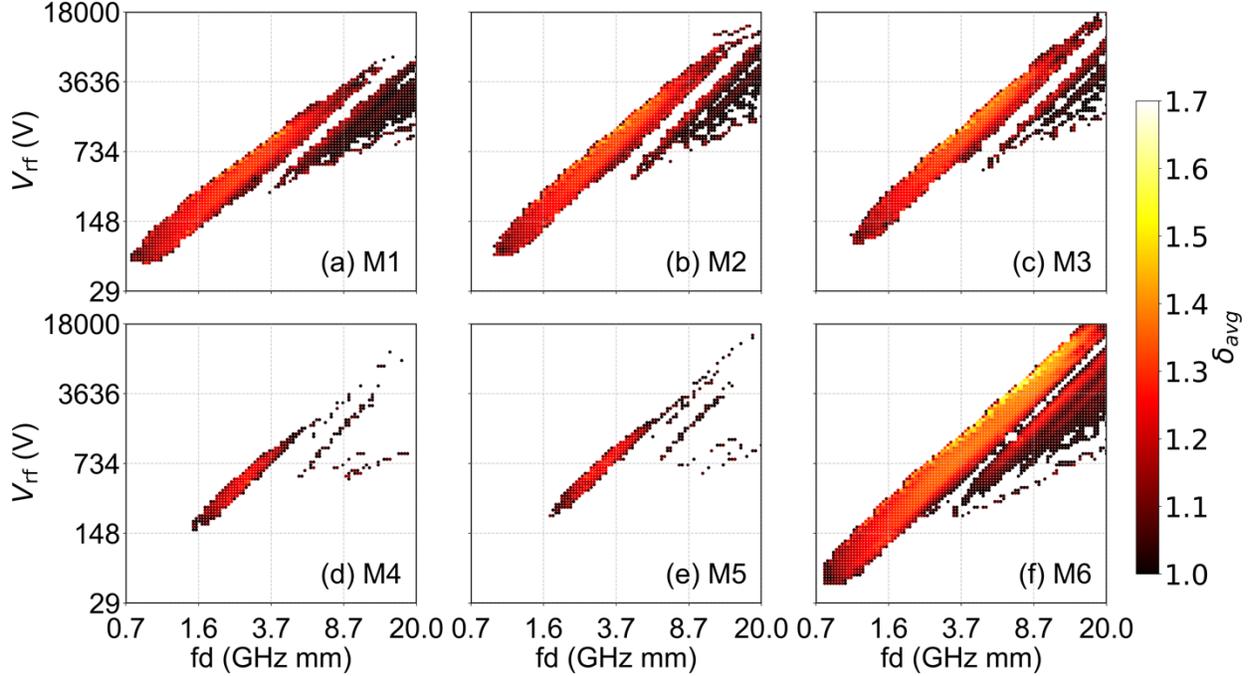

Fig. 1: Multipactor susceptibility charts for all six material property sets (Table 1) generated using interpolation following 3D electromagnetic PIC simulations (as described in Section 2.1) using the same simulation parameters as in Iqbal et al., 2022. The color bar represents the time-averaged electron growth rate $\delta_{avg}$ across the ($V_{rf}$, fd) plane. Colored regions where $\delta_{avg} > 1$ denote susceptibility to multipactor growth, whereas white areas with $\delta_{avg} < 1$ correspond to stable, multipactor-free operation.

*2.2 Machine Learning Workflow Overview*

The workflow followed in this study is illustrated in Fig. 2. It outlines the dataset generation, feature analysis, model selection, training strategy, evaluation metrics, and models used to predict multipactor susceptibility for different materials. Each component of the pipeline is described in more detail in the subsequent sections.

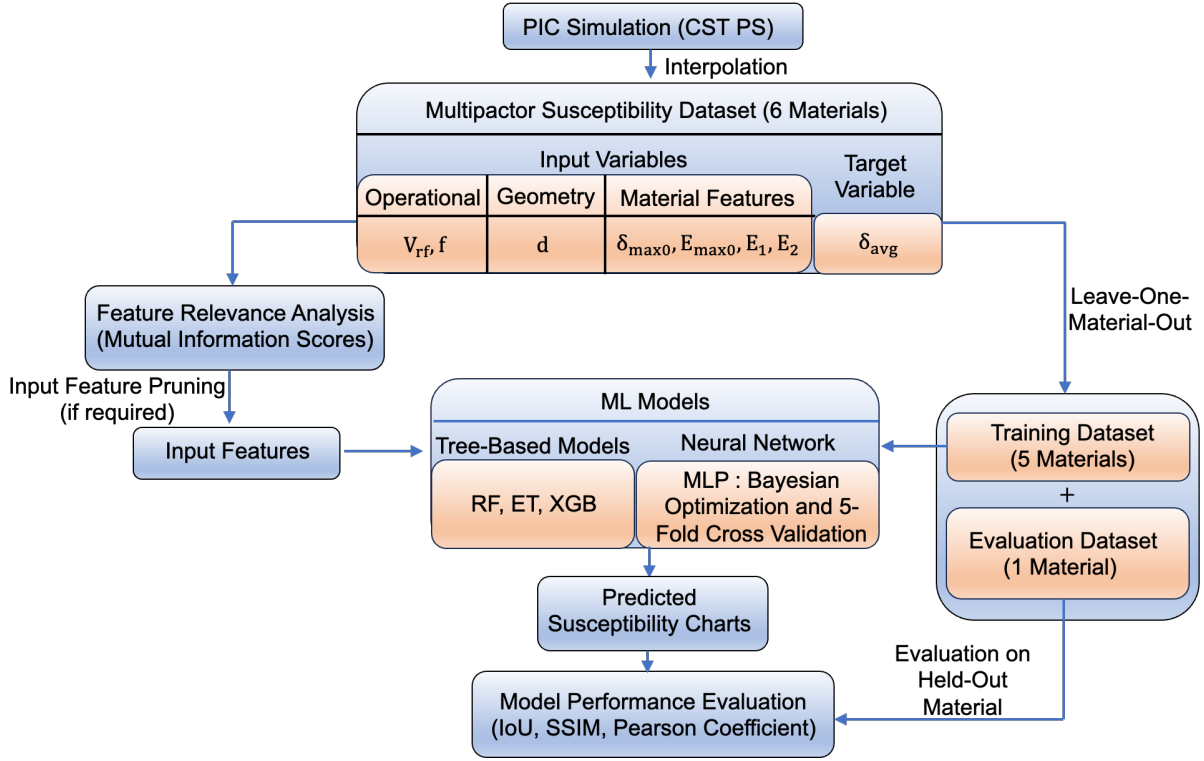

Fig. 2. Workflow of the supervised machine learning framework for multipactor susceptibility prediction. The pipeline begins with PIC simulations in CST Particle Studio to generate a multipactor susceptibility dataset for six materials. Input features include operational and geometric variables ($V_{rf}$, $f$, $d$), and material-dependent secondary electron yield (SEY) properties ($\delta_{max0}$, $E_{max0}$, $E_1$, $E_2$), while the target is the time-averaged ensemble electron growth rate $\delta_{avg}$. Feature relevance is first assessed via mutual information (MI) scores, with potential pruning of low-importance features. Models are trained using a leave-one-material-out strategy: five materials are used for training and optimization (via 5-fold cross-validation for ANN), and the sixth is held out for evaluation. Multiple regression models are employed—Random Forest (RF), Extra Trees (ET), and XGBoost (XGB)—along with a funnel-structured Multilayer Perceptron (MLP) with Bayesian optimization. Performance is evaluated using IoU, SSIM, and Pearson correlation metrics.

## 2.3 Target Susceptibility Metric

The primary target variable for machine learning prediction in this study is the time-averaged electron growth rate ($\delta_{avg}$), computed from the PIC simulations as described in Section 2.1. This susceptibility parameter provides a quantitative measure of multipactor strength under specified operational and material conditions. For supervised learning tasks, the models were trained to regress the continuous value of $\delta_{avg}$, enabling prediction of multipactor susceptibility across the operational parameter space.

While this problem can also be formulated as a classification task, a regression-based modeling approach was adopted. Predicting $\delta_{avg}$ as a continuous variable enables the model to learn underlying physical trends associated with multipactor growth, rather than simply delineating susceptible and non-susceptible regions. By contrast, a classification approach (e.g., predicting whether $\delta_{avg} > 1$) would reduce the task to image recognition over susceptibility charts. Such a formulation would not necessarily promote physical interpretability and would be severely constrained by the limited dataset size, which consists of only six susceptibility charts. A regression-oriented framework is thus better suited to leveraging the available data and preserving the physical interpretability of model predictions.

*2.4 Input Features*

We consider the multipactor strength, quantified by the time averaged electron growth rate $\delta_{avg}$, to be a function of the geometric and operational parameters, as well as the intrinsic material responses governed by secondary electron yield (SEY) characteristics. This relationship can be expressed as:

$$\delta_{avg} = g(\text{Geometry, Operational conditions, Intrinsic SEY characteristics}) \quad (1)$$

The geometric and operational features include the RF voltage amplitude ($V_{rf}$) and the frequency-gap distance product (fd), following the analytical similarity scaling introduced by Vaughan (1988). The intrinsic SEY properties of a material are characterized by four parameters from Vaughan's SEY model (Vaughan, 1989, 1993) as described in Section 2.1. Consequently, the susceptibility parameter is modeled as:

$$\delta_{avg} = g(V_{rf}, \text{fd}, \delta_{max0}, E_{max0}, E_1, E_2) \quad (2)$$

The selection of these input features is motivated by a range of previous experimental, analytical, and simulation studies that established their critical roles in determining multipactor onset and growth (Iqbal et al., 2023a; Kishek et al., 1998; Vaughan, 1988).

*2.5 Feature Relevance Analysis*

To assess the relative importance of each input feature in predicting the multipactor susceptibility, mutual information (MI) scores ("Entropy, Relative Entropy, and Mutual Information," 2005; Pedregosa et al., 2011) were computed between each input feature and the target variable $\delta_{avg}$. Mutual Information (MI) is a model-independent measure that quantifies how much information one variable provides about another, capturing both linear and nonlinear relationships. This makes it a robust tool for identifying which input features are most informative for predicting the target variable. The mutual information (MI) between a feature X and the target variable Y is defined as:

$$\text{MI}(X; Y) = \iint p(x, y) \log\left(\frac{p(x, y)}{p(x)p(y)}\right) dxdy. \tag{3}$$

where $p(x, y)$ is the joint probability density function, and $p(x)$ and $p(y)$ are the marginal densities. The symbols x and y denote specific realizations of random variables X and Y, respectively. The joint and marginal densities $p(x, y), p(x), p(y)$ are evaluated at these values and integrated over their respective domains. In this study, MI was estimated using the *mutual_info_regression* function from the scikit-learn library, which applies a nonparametric k-nearest neighbors approach ($k = 5$ used for this study) to approximate the densities. The computation was averaged across 100 independent runs to account for estimator randomness. Final scores were normalized to sum to 1 (Fig. 3) to highlight their relative contribution to predicting $\delta_{\text{avg}}$.

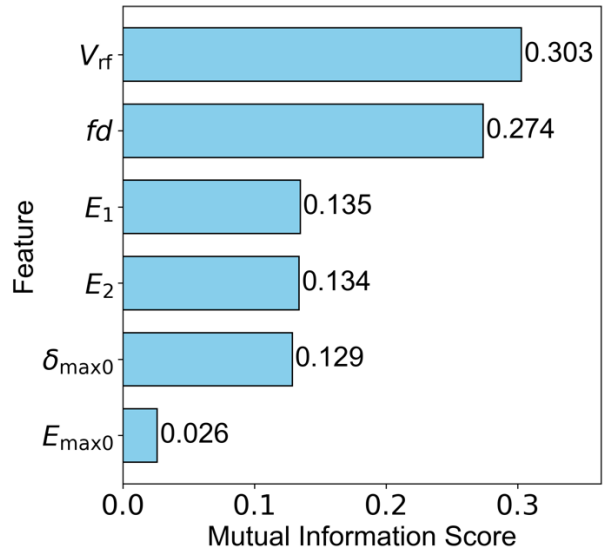

Fig. 3. Normalized mutual information (MI) scores indicating the relative contribution of each input feature to predicting the electron growth rate $\delta_{\text{avg}}$. Scores are normalized by the total mutual information across all features such that the bars sum to 1, highlighting each feature's relative importance in predicting $\delta_{\text{avg}}$.

Fig. 3 presents the relative mutual information (MI) scores of the six input features used in this study, providing insight into their relative predictive contributions to multipactor susceptibility. The RF voltage amplitude ($V_{rf}$) and the frequency-gap distance product ($fd$) exhibit the highest MI scores, reflecting their dominant role in governing electron growth and susceptibility onset. Among the material-dependent secondary electron yield (SEY) features, the first ($E_1$) and second ($E_2$) crossover energies, along with the maximum SEY ($\delta_{max0}$), contribute moderate but meaningful information to the model.

In contrast, the incident energy at maximum SEY ($E_{max0}$) exhibits a significantly lower MI score, suggesting that it contributes minimally to the model's predictive capability. This finding is physically consistent with prior work— Rosario and Edén (2012) showed that the multipactor threshold is directly governed by the first and second crossover energies ($E_1$ and $E_2$), where the SEY curve intersects unity. While $E_{max0}$ defines the peak location and overall shape of the SEY curve (Vaughan, 1989)—and thereby influences the positions of $E_1$ and $E_2$—its effect on multipactor susceptibility is indirect, mediated through these crossover energies. Therefore, the low MI score of $E_{max0}$ aligns with its secondary role in influencing multipactor susceptibility, especially when $E_1$ and $E_2$ are already included in the input features list.

While an alternative approach of feature relevance analysis could involve systematically removing each input feature and evaluating the resulting model performance, this brute-force pruning strategy was not pursued. Given the small number of input features, the strong prior physical understanding of their relevance, and the potential for introducing model-specific biases through retraining, we instead adopted a model-independent, non-parametric mutual information (MI)-based analysis. This approach enables efficient and interpretable evaluation consistent with the established physics of multipactor susceptibility.

Based on this preliminary analysis, further investigation was conducted to validate whether $E_{max0}$ could be considered a redundant feature. Model performance comparisons with and without $E_{max0}$ as an input feature are presented in Section 3 to validate this assessment.

*2.6 Machine Learning Models*

In this study, various machine learning models—including linear regression models, tree-based nonlinear ensemble models, and neural network models—were explored to predict the multipactor susceptibility parameter $\delta_{avg}$ based on the selected input features.

*2.6.1 Linear Regression Models*

To assess the suitability of linear models, several regression techniques were tested, including Linear Regression, Ridge, Lasso, ElasticNet, and Bayesian Ridge. However, all failed to predict multipactor susceptibility, consistently predicting $\delta_{avg} < 1$ across the entire $(V_{rf}, fd)$ space. This outcome reflects the models' inability to capture the nonlinear dependencies between the input features and the target variable. Consequently, linear models were not pursued further, and the focus shifted to nonlinear approaches.

Table 2: Performance of linear regression models in predicting multipactor susceptibility.

| Model | Multipactor Prediction |
|---|---|
| Linear Regression | $\delta_{avg} < 1$ for all $(V_{rf}, fd)$ |
| Ridge Regression | $\delta_{avg} < 1$ for all $(V_{rf}, fd)$ |
| Lasso Regression | $\delta_{avg} < 1$ for all $(V_{rf}, fd)$ |
| ElasticNet Regression | $\delta_{avg} < 1$ for all $(V_{rf}, fd)$ |
| Bayesian Ridge Regression | $\delta_{avg} < 1$ for all $(V_{rf}, fd)$ |

*2.6.2 Tree-Based Ensemble Models*

Tree-based ensemble models are well-suited for modeling complex, nonlinear relationships without requiring extensive feature engineering. Two bagging-based (Breiman, 1996) models were employed:

- Random Forest (RF) regression (Breiman, 2001): A bootstrap-aggregated ensemble of decision trees, trained with random feature selection at each split to reduce overfitting and variance.
- Extra Trees (ET) regression (Geurts et al., 2006): An ensemble similar to RF but introducing additional randomness by selecting split thresholds at random for each feature, further decorrelating the trees and enhancing robustness.

In addition to these bagging methods, a boosting-based (Schapire, 1999) ensemble model—Extreme Gradient Boosting (XGBoost) (Chen and Guestrin, 2016)—was also explored. XGBoost is a scalable, regularized boosting algorithm that sequentially fits new decision trees to correct the residual errors of previous trees.

*2.6.3 Neural Network: Multilayer Perceptrons (MLP)*

A Multilayer Perceptron (MLP) consists of an input layer, one or more hidden layers, and an output layer. Each layer $l$ transforms its input $X^{(l-1)}$ through a linear transformation followed by a nonlinear activation function:

$$\mathbf{X}^{(l)} = \boldsymbol{\sigma}(\mathbf{W}^{(l)}\mathbf{X}^{(l-1)} + \mathbf{b}^{(l)}) \tag{4}$$

where $\mathbf{W}^{(l)}$ and $\mathbf{b}^{(l)}$ are the weight matrix and bias vector for layer $l$ respectively, and $\boldsymbol{\sigma}(\cdot)$ denotes the nonlinear activation function (e.g., ReLU or tanh).

In this study, an MLP model was implemented in TensorFlow/Keras to approximate the mapping from input features to the target variable $\delta_{\text{avg}}$. The network architecture denoted $F(\theta_h)$, followed a funnel structure, in which the number of neurons ($N_{\text{neurons}}$) approximately halved from one hidden layer to the next:

$$N_{\text{neurons}}^{(l+1)} \approx 0.5 \times N_{\text{neurons}}^{(l)}. \tag{5}$$

This structure encourages efficient compression of feature representations as the data propagates through the network. The tunable hyperparameters $\theta_h$ included:

- number of layers L: sampled from $\{1,2,\ldots,10\}$
- initial neurons count ($n_0$) for the first hidden layer: sampled from $[50, 1024]$
- activation function ($\sigma$): either ReLU or tanh
- L2 regularization coefficient ($\alpha$): sampled logarithmically between $10^{-6}$ and $10^{-2}$
- learning rate ($\eta$): sampled logarithmically between $10^{-4}$ and $10^{-1}$

Hyperparameter optimization was performed using Bayesian optimization (Shahriari et al., 2016) for each material separately. The corresponding dataset was excluded (leave-one-material-out), and the model was trained and validated on the remaining five materials. A 5-fold cross-validation (Kohavi, 1995) was conducted within the training data to evaluate each candidate architecture's generalization performance. Table 3 summarizes the hyperparameter search space and training configuration used for MLP model development. The hyperparameters were optimized via Bayesian optimization using 5-fold cross-validation on the training set, with the search space designed to explore both depth and regularization settings.

Table 3: Hyperparameter search space and training configuration for MLP models

| Item | Choice |
|---|---|
| Optimizer | Bayesian Optimization |
| Model | MLPs implemented using Keras/Tensorflow |
| Number of Layers | 1 to 10 |
| Funnel Architecture | Yes (dynamic) |
| Neurons in First Layer | 50 to 1024 |

| Item | Choice |
|---|---|
| Activation Function | 'relu' or 'tanh' |
| $\alpha$ | log-uniform $[10^{-6}, 10^{-2}]$ |
| Learning Rate | log-uniform $[10^{-5}, 10^{-1}]$ |
| Optimization Iterations | 30 |
| Loss Function | IoU, SSIM, scalarized loss function |
| Validation Split Method | 5-Fold Cross-Validation |

The MLP models were trained using three different objective functions:

- Negative mean IoU over 5 folds: $\Gamma(\theta_h) = -\text{IoU}_{CV}(\theta_h)$
- Negative mean SSIM over 5 folds: $\Gamma(\theta_h) = -\text{SSIM}_{CV}(\theta_h)$
- A scalarized loss combining both metrics: $\Gamma(\theta_h) = -(0.5 \times \text{IoU}_{CV}(\theta_h) + 0.5 \times \text{SSIM}_{CV}(\theta_h))$

The evaluation metrics Intersection over Union (IoU) and Structural Similarity Index Measure (SSIM) are defined in detail in Section 2.8.

Once the optimal hyperparameters were identified, the final model for each material was retrained on the full 5-material dataset using mean squared error (MSE) loss and optimized using the Adam algorithm.

*2.6.4. Rationale for Model Selection*

The machine learning models adopted in this study were selected for their ability to capture the nonlinear relationships between multipactor susceptibility ($\delta_{avg}$) and the combined geometric, operational, and material-dependent features. Due to the complex and clustered nature of the susceptibility data across different materials, model interpretability, robustness to data sparsity, and nonlinear expressiveness were key selection criteria.

Tree-based ensemble methods—Random Forest (RF), Extra Trees (ET), and Extreme Gradient Boosting (XGB)—were employed for their ability to model nonlinear interactions without requiring prior feature transformation. RF and ET use bagging strategies that reduce variance by aggregating predictions from multiple randomized trees trained in parallel. In contrast, XGB employs a boosting strategy, fitting trees sequentially to correct residual errors. Although this approach carries a higher risk of overfitting, it can in some cases achieve superior predictive accuracy.

The Multilayer Perceptron (MLP) architecture implemented in TensorFlow/Keras was designed to efficiently capture complex feature interactions through a funnel structure and Bayesian hyperparameter optimization. To mitigate overfitting—especially critical in small datasets—L2 regularization and 5-fold cross-validation were employed.

The combination of tree-based ensembles and MLP enables comparative insights across algorithmic classes, while grounding the modeling approach in both interpretability and flexibility. Their relative performance is discussed in detail in Section 3.

*2.7 Model Evaluation and Cross-Validation*

The model evaluation strategy in this study was designed to reflect real-world scenarios where new materials are introduced in RF devices and accelerator components. To assess cross-material generalization, a leave-one-chart-out cross-validation approach was adopted. Each susceptibility chart—corresponding to a distinct material property set—was treated as an independent group. In each iteration, the model was trained on data from five materials and evaluated on the held out sixth, enabling a systematic assessment of the model's ability to generalize to previously unseen material properties.

During Bayesian optimization of the neural network architectures, a 5-fold cross-validation strategy was performed within the training subset (i.e., the five charts used in a given iteration) to guide hyperparameter selection based on evaluation metrics discussed in Section 2.8. The held-out chart was excluded from all stages of hyperparameter tuning. Once the optimal architecture was identified, the final model was retrained on the entire five-material dataset and was tested on the unseen chart using the same leave-one-chart-out protocol.

*2.8 Evaluation Metrics*

To guide the machine learning models' training and hyperparameter tuning (for the MLP) and to assess their performance, several evaluation metrics were employed. In high-power RF and accelerator systems, it is critical to avoid operating conditions that lie within multipactor susceptibility regions in the ($V_{rf}$, fd) space. Therefore, beyond predicting the electron growth rate $\delta_{avg}$ accurately, it is equally—if not more—important to precisely delineate the boundaries of susceptible regions. Accordingly, the selected metrics were designed to assess not only numerical accuracy but also the spatial and structural fidelity of the predicted susceptibility maps relative to the simulation-derived reference charts.

*2.8.1 Mean Squared Error (MSE):*

The Mean Squared Error (MSE) is a standard loss function for regression tasks that measures the average of the squared differences between the predicted and true values. It is defined as:

$$\text{MSE} = \frac{1}{n}\sum_{i=1}^{n}(y_i - \hat{y}_i)^2 \tag{6}$$

where $y_i$ is the true value, $\hat{y}_i$ is the predicted value, and $n$ is the total number of samples. In this study, while MSE was minimized during model training, model performance was evaluated using metrics better suited for capturing spatial overlap and structural fidelity —namely, Intersection over Union (IoU), Structural Similarity Index (SSIM), and Pearson Correlation Coefficient.

*2.8.2 Intersection over Union (IoU):*

Intersection over Union (IoU) measures the degree of overlap between the predicted and reference susceptible regions. It is defined as:

$$\text{IoU} = \frac{|P \cap S|}{|P \cup S|} \tag{7}$$

where $P$ is the predicted multipactor susceptibility region (corresponding to $\delta_{\text{avg,predicted}} \geq 1$) and $S$ is the simulated ground truth multipactor susceptibility region (corresponding to $\delta_{\text{avg,simulated}} \geq 1$) obtained from Particle-in-Cell (PIC) simulations. IoU evaluates how well the model identifies susceptible regions, providing a physically meaningful measure of prediction accuracy.

*2.8.3 Structural Similarity Index Measure (SSIM):*

The Structural Similarity Index (SSIM) assesses the perceptual similarity between the predicted and reference susceptibility charts by comparing their luminance, contrast, and structural information. It is computed as:

$$\text{SSIM}(x, y) = \frac{(2\mu_x\mu_y + C_1)(2\sigma_{xy} + C_2)}{(\mu_x^2 + \mu_y^2 + C_1)(\sigma_x^2 + \sigma_y^2 + C_2)} \tag{8}$$

where $\mu_x$ and $\mu_y$ are the means of $x$ and $y$, $\sigma_x^2$ and $\sigma_y^2$ are their variances, $\sigma_{xy}$ is the covariance, and $C_1$ and $C_2$ are small constants for numerical stability. A higher SSIM value indicates greater structural resemblance between predicted and true susceptibility distributions.

*2.8.4 Pearson Correlation Coefficient:*

The Pearson correlation coefficient measures the linear relationship between the predicted and true $\delta_{avg}$ values across the dataset, quantifying how well the predicted values follow the trend of the true values. It is defined as:

$$r = \frac{\sum_{i=1}^{n}(y_i - \bar{y})(\hat{y}_i - \bar{\hat{y}})}{\sqrt{\sum_{i=1}^{n}(y_i - \bar{y})^2} \sqrt{\sum_{i=1}^{n}(\hat{y}_i - \bar{\hat{y}})^2}} \tag{9}$$

where $\bar{y}$ and $\bar{\hat{y}}$ are the means of the true and predicted values, respectively. A Pearson coefficient close to 1 indicates strong agreement in trends between the predicted and true $\delta_{avg}$ values.

## 3. Results and Discussion

This section presents the results of machine learning model performance in predicting multipactor susceptibility. We begin by evaluating RF model predictions and benchmarking against Monte Carlo simulations. We also evaluate the impact of including $E_{max0}$ as an input feature. Subsequently, we present susceptibility prediction results obtained using different machine learning models — Extra Trees (ET), Extreme Gradient Boosting (XGB), and Multi-Layer Perceptron (MLP) — and visually compare the predicted susceptibility charts. A quantitative comparison of model performance is conducted across multiple evaluation metrics, providing insights into the relative strengths and limitations of each modeling approach.

*3.1 Random Forest (RF) model predictions, feature pruning validation, and benchmarking against Monte Carlo (MC) simulations*

Building on the analysis in Section 2.5, to assess the role of the incident energy at maximum secondary electron yield $E_{max0}$, Random Forest (RF) regression models were trained and evaluated both with and without including $E_{max0}$ as an input feature.

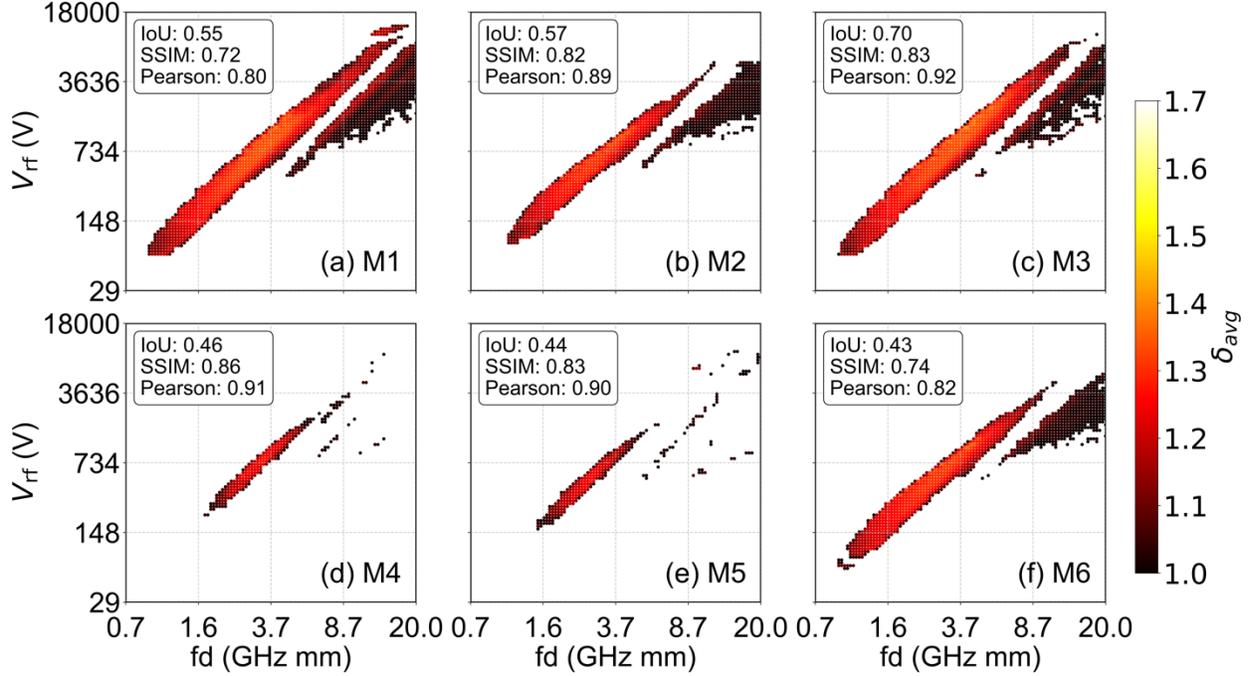

Fig. 4: Predicted multipactor susceptibility charts in the $(V_{rf}, fd)$ plane for six material property sets (M1–M6) using Random Forest (RF) regression models trained with all input features, including $E_{max0}$. Electron growth rate $\delta_{avg}$ is shown in the color bar. Regions with $\delta_{avg} > 1$ correspond to multipactor susceptibility.

Fig. 4(a)–(f) shows the predicted susceptibility charts for the six material property sets (M1–M6) using RF models trained with the full set of six input features, including $E_{max0}$. Each subplot corresponds to the held-out material in a leave-one-chart-out cross-validation, thus evaluating the model's ability to generalize to previously unseen material properties.

The predicted susceptibility regions, defined by $\delta_{avg} > 1$, closely follow the qualitative shape and extent of the susceptibility zones observed in the ground truth charts from PIC simulations (see Fig. 1). The color gradients across the $(V_{rf}, fd)$ domain reflect a smooth and physically reasonable variation in $\delta_{avg}$, with high-growth regions appearing in locations consistent with the multipactor physics and PIC simulations. This consistency indicates that the model effectively learned the influence of both operational and material parameters.

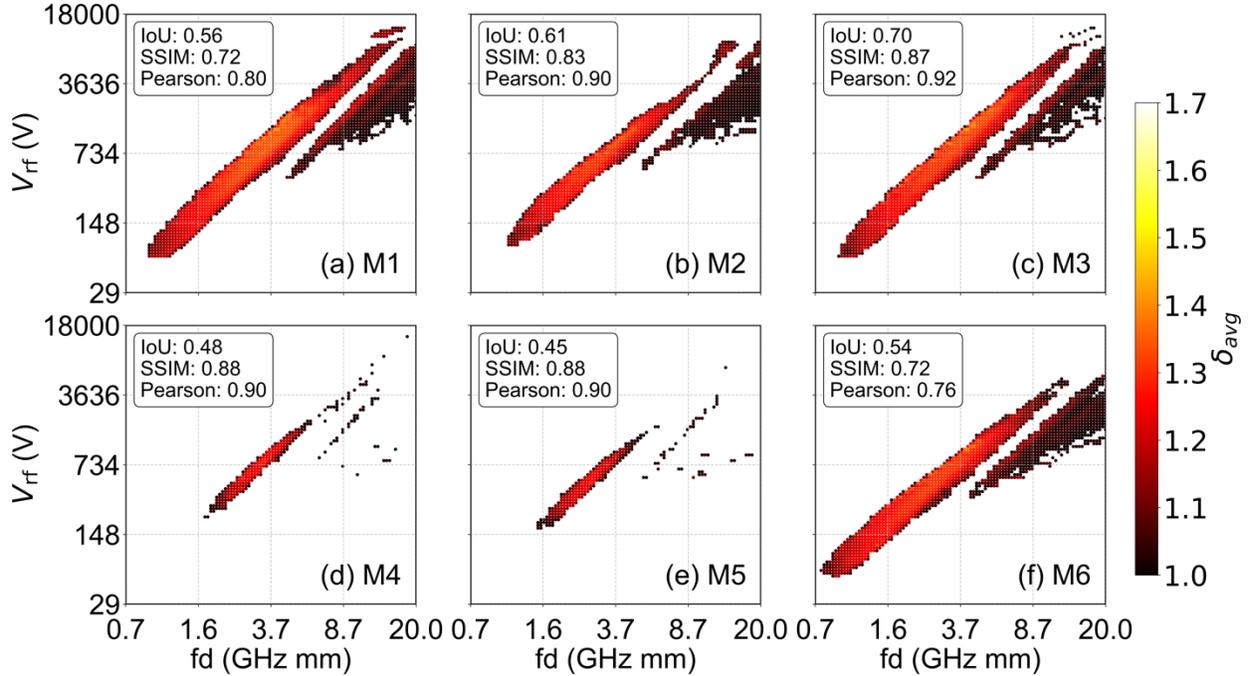

Fig. 5: Predicted multipactor susceptibility charts for six material property sets (M1–M6) using RF regression models trained without $E_{max0}$ as an input feature. Comparison with Fig. 4 shows that exclusion of $E_{max0}$ results in minimal visual effect on susceptibility predictions, supporting the redundancy of this feature.

To evaluate the potential redundancy of $E_{max0}$, the same RF regression models were retrained after removing $E_{max0}$ from the input feature set. The corresponding predicted susceptibility charts are shown in Fig. 5(a)–(f). The predicted susceptibility regions without $E_{max0}$ appear largely similar to those obtained with the full input set, suggesting that the exclusion of $E_{max0}$ does not degrade model performance.

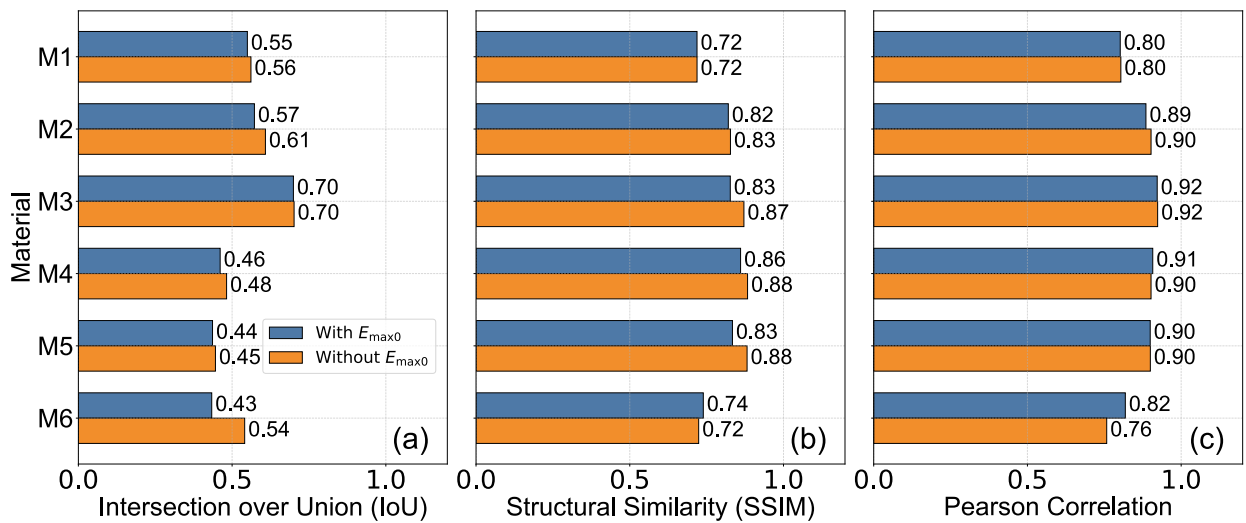

Fig. 6: Comparison of evaluation metrics (a) Intersection over Union (IoU), (b) Structural Similarity Index (SSIM), and (c) Pearson Correlation Coefficient for RF model predictions with (blue bars) and without (orange bars) $E_{max0}$ as an input feature for each material property set. Slight improvements or negligible changes in scores confirm that $E_{max0}$ can be safely pruned without loss of predictive accuracy.

This observation is quantitatively supported by the evaluation metrics summarized in Fig. 6, which compares Intersection over Union (IoU) (Fig. 6(a)), Structural Similarity Index (SSIM) (Fig. 6(b)), and Pearson Correlation Coefficient (Fig. 6(c)) scores for each material, with (blue bars) and without (orange bars) $E_{max0}$ as an input feature. Across all six materials, the metrics remain largely stable, with slight improvements observed in some cases upon removing $E_{max0}$. This trend is consistent with the earlier mutual information analysis of Section 2.5, which indicated that $E_{max0}$ had minimal predictive contribution relative to other features. The slight improvements after removing $E_{max0}$ from the input feature set can be attributed to a reduction in feature noise or redundancy—suggesting that the inclusion of $E_{max0}$ despite its limited direct relevance to multipactor onset, may have introduced statistical noise with limited predictive value that interfered with model training. Based on these results, $E_{max0}$ was pruned from the input feature set for all subsequent modeling tasks, improving model simplicity without compromising prediction fidelity.

As discussed in Section 2.1, comprehensive multipactor susceptibility datasets are scarce in the literature. Nonetheless, we were able to benchmark the RF model's predictions against results from a contemporary approach, i.e., a two-dimensional (2D) Monte Carlo (MC) simulation for copper, which corresponds to Material M2 in our dataset. From the published susceptibility chart in Fig. 6 of Iqbal et al. (2022), we computed the Intersection over Union (IoU), Structural Similarity Index (SSIM), and Pearson coefficient scores between the 2D MC and 3D PIC results, obtaining values of 0.65, 0.94, and 0.95, respectively. These metrics are comparable to the RF model's performance on M2 when $E_{max0}$ is excluded as an input feature (IoU = 0.61, SSIM = 0.83, Pearson Coefficient = 0.90), suggesting that the machine learning model achieves predictive accuracy comparable to computationally intensive MC simulations. Given that this level of accuracy was obtained using a limited training set of only five materials, the results strongly support the promise of ML-based approaches and motivate further investment in data generation and model development for multipactor prediction.

*3.2 Performance of Tree-Based Models*

Figs. 7 and 8 present the predicted multipactor susceptibility charts for all six materials using the Extra Trees (ET) and Extreme Gradient Boosting (XGB) models, respectively. In each plot, the spatial distribution of the predicted electron growth rate $\delta_{avg}$ is shown across the ($V_{rf}$, fd) plane, along with the corresponding IoU, SSIM, and Pearson Correlation Coefficient metrics.

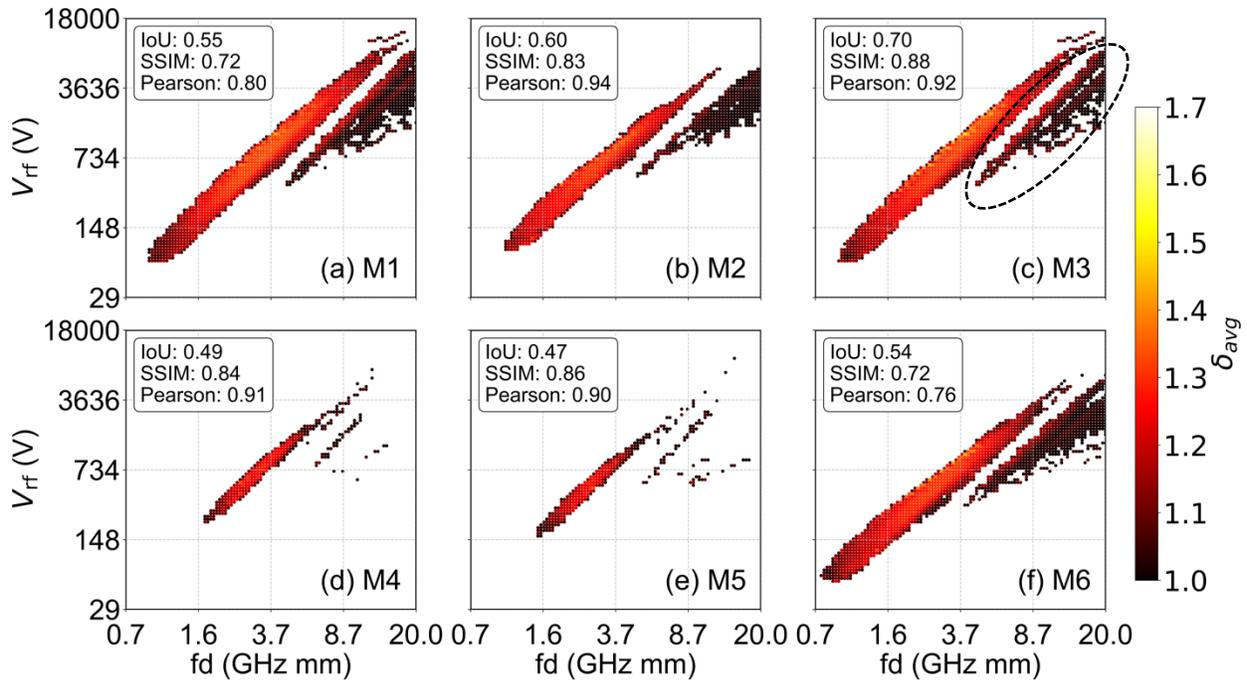

Fig. 7: Predicted multipactor susceptibility charts for six material property sets (M1–M6) using Extra Trees (ET) regression models.

As seen in Fig. 7, the ET model demonstrates strong predictive fidelity across all materials, achieving performance comparable to the Random Forest (RF) baseline (Fig. 5), with consistently high evaluation metrics scores. Notably, for material M3, the ET model appears to better resolve the higher order susceptibility bands (outlined by the dotted elliptical region Fig. 7(c)), suggesting improved sensitivity to subtle variations in the $\delta_{avg}$ landscape compared to the RF model.

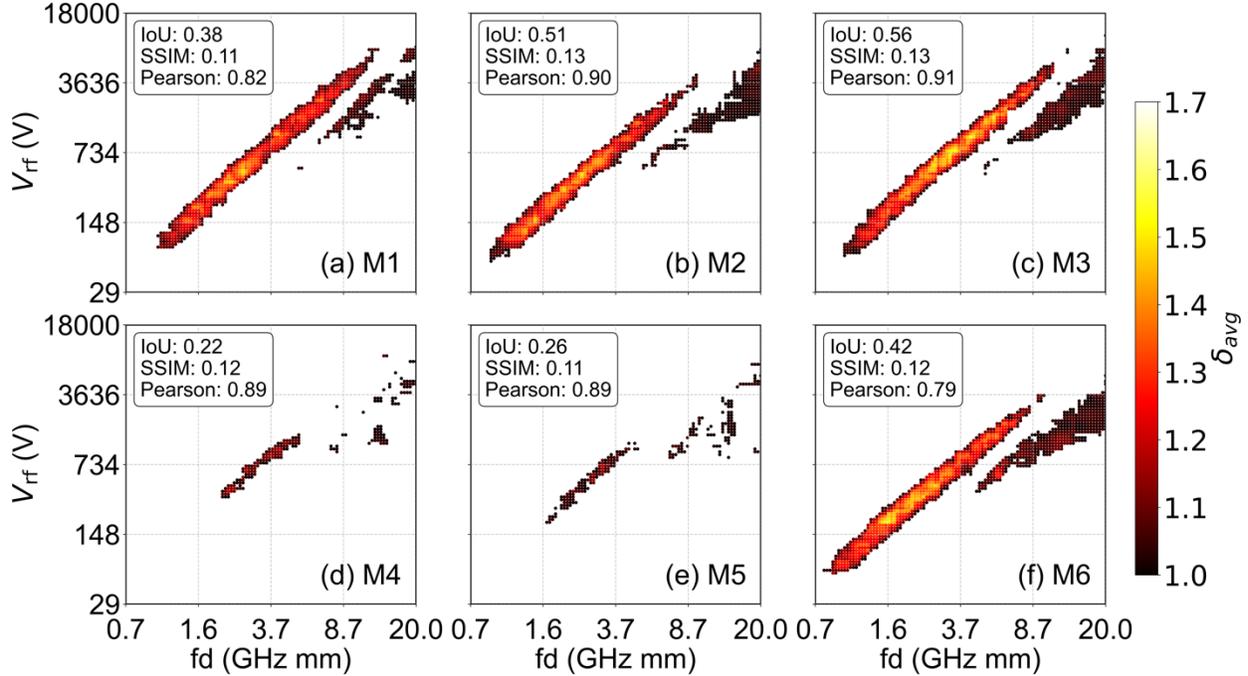

Fig. 8: Predicted multipactor susceptibility charts for six material property sets (M1–M6) using Extreme Gradient Boosting (XGBoost) models.

In contrast, although the XGB model (Fig. 8) captures the general trend of susceptibility onset, it exhibits a notable decline in structural similarity, as reflected by lower SSIM values. This suggests that while XGB is capable of identifying the broad susceptibility growth, it struggles to accurately reconstruct the spatial organization of the susceptibility region. This limitation can be attributed to XGB's sequential boosting strategy, which excels at capturing localized, complex variations but is more prone to overfitting in problems characterized by globally smooth trends. In comparison, the bagging strategies employed by RF and ET build multiple decorrelated decision trees in parallel and aggregate their outputs, producing smoother and more stable approximations and, consequently, superior structural fidelity and overall predictive performance.

*3.3 Performance of Neural Network Models (MLPs)*

*3.3.1 MLP Optimized for IoU:*

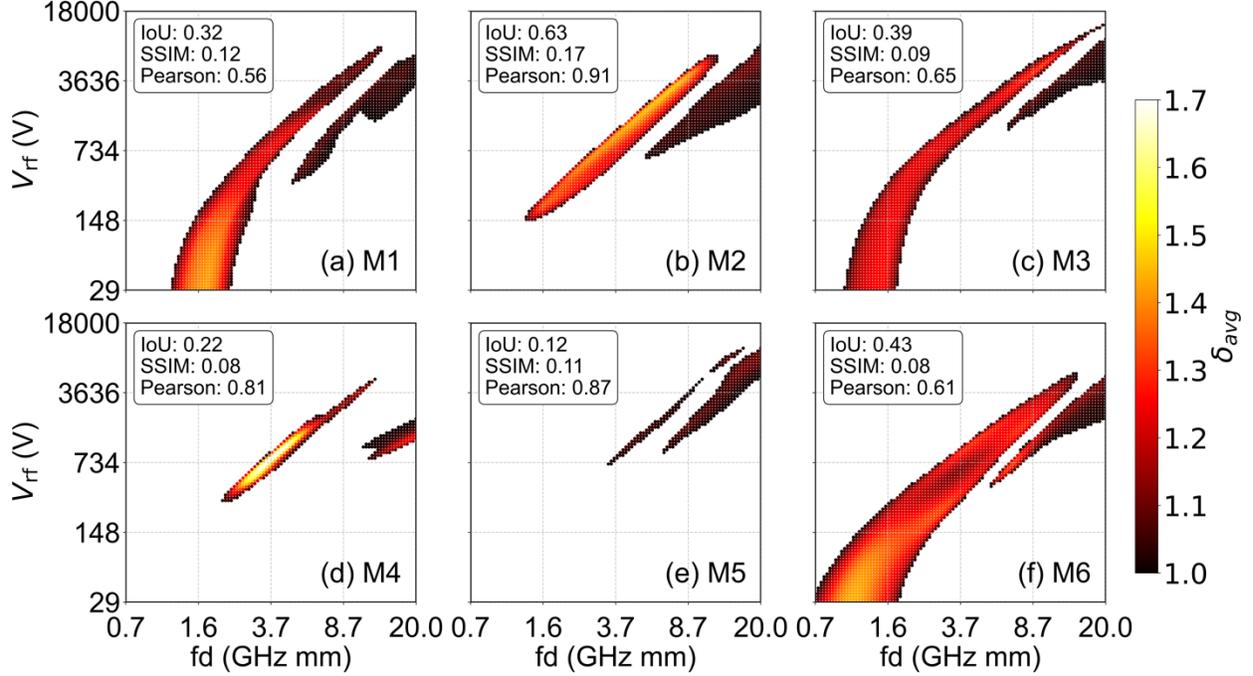

Fig. 9: Predicted multipactor susceptibility charts for six material property sets (M1–M6) using MLP models optimized for IoU as the sole objective function ($\Gamma(\theta_h) = -\text{IoU}_{CV}(\theta_h)$).

When the MLP was trained using Intersection over Union (IoU) as the sole objective function ($\Gamma(\theta_h) = -\text{IoU}_{CV}(\theta_h)$), the predicted susceptibility charts exhibited poor agreement with the reference data across most materials. Although the optimization process yielded high cross-validation IoU scores during training (~0.71–0.74), the final model's generalization performance on the held-out material was inconsistent and weak. In particular, susceptibility boundaries were inaccurately predicted or often fragmented, and SSIM values remained extremely low (≤ 0.17), indicating little to no structural fidelity. While some materials such as M2 showed moderate IoU and Pearson correlation (0.63 and 0.91, respectively), overall, the IoU-only objective led to brittle predictions that overfit susceptibility "blobs" without learning meaningful structure. Performance was substantially worse than that of tree-based ensemble models (e.g., RF, ET), revealing the limitations of optimizing purely for spatial overlap in the multipactor context.

*3.3.2 MLP Optimized for SSIM:*

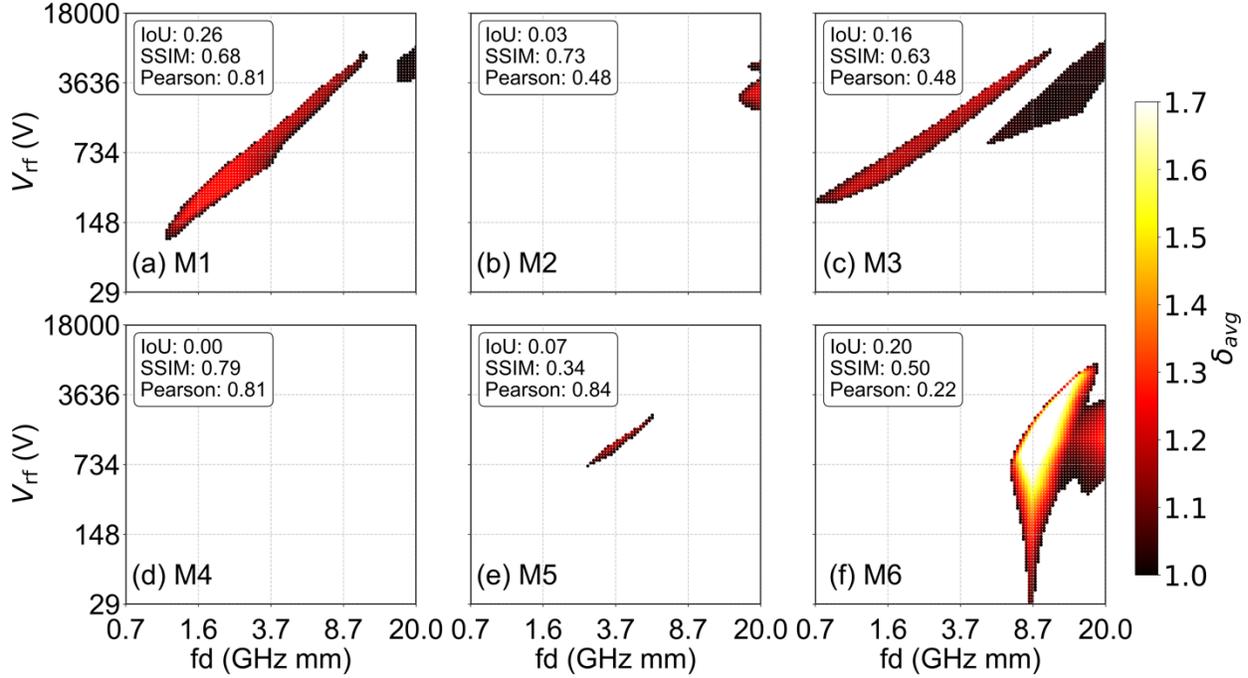

Fig. 10: Predicted multipactor susceptibility charts for six material property sets (M1–M6) using MLP models optimized for SSIM as the sole objective function ($\Gamma(\theta_h) = -\text{SSIM}_{CV}(\theta_h)$).

Using SSIM as the sole training objective ($\Gamma(\theta_h) = -\text{SSIM}_{CV}(\theta_h)$) led to a different failure mode. Although the predicted susceptibility regions exhibited some localized structural resemblance and relatively high SSIM values (up to 0.84 for M5), they were poorly placed or missing altogether in several cases (e.g., M4, M5). IoU scores were extremely low across all materials—several near zero—indicating a severe mismatch in the predicted versus actual regions of multipactor growth. This outcome reflects the inability of SSIM alone to enforce correct localization, performing even worse than the IoU-optimized models in terms of actionable susceptibility detection.

*3.3.3 MLP Optimized for Scalarized IoU + SSIM:*

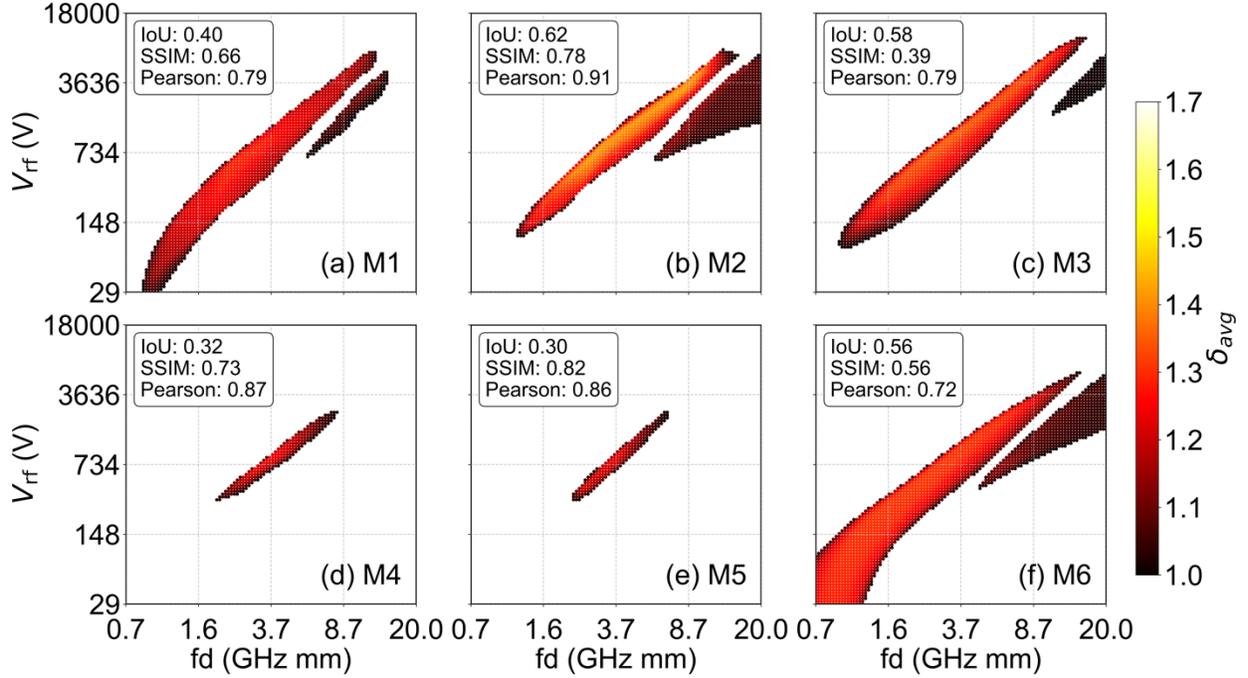

Fig. 11: Predicted multipactor susceptibility charts for six material property sets (M1–M6) using MLP models optimized for a scalarized loss combining IoU and SSIM ($\Gamma(\theta_h) = -(0.5 \times \text{IoU}_{CV}(\theta_h) + 0.5 \times \text{SSIM}_{CV}(\theta_h))$)

In contrast to sole loss functions, MLP models trained with a scalarized loss function combining IoU and SSIM ($\Gamma(\theta_h) = -(0.5 \times \text{IoU}_{CV}(\theta_h) + 0.5 \times \text{SSIM}_{CV}(\theta_h))$) demonstrated significantly improved predictive performance (Fig. 11). Across all six materials, these models achieved a better balance between regional overlap and shape consistency, with IoU values ranging from 0.30 (for material M5) to 0.62 (for material M2) and SSIM values reaching up to 0.82 (for material M5).

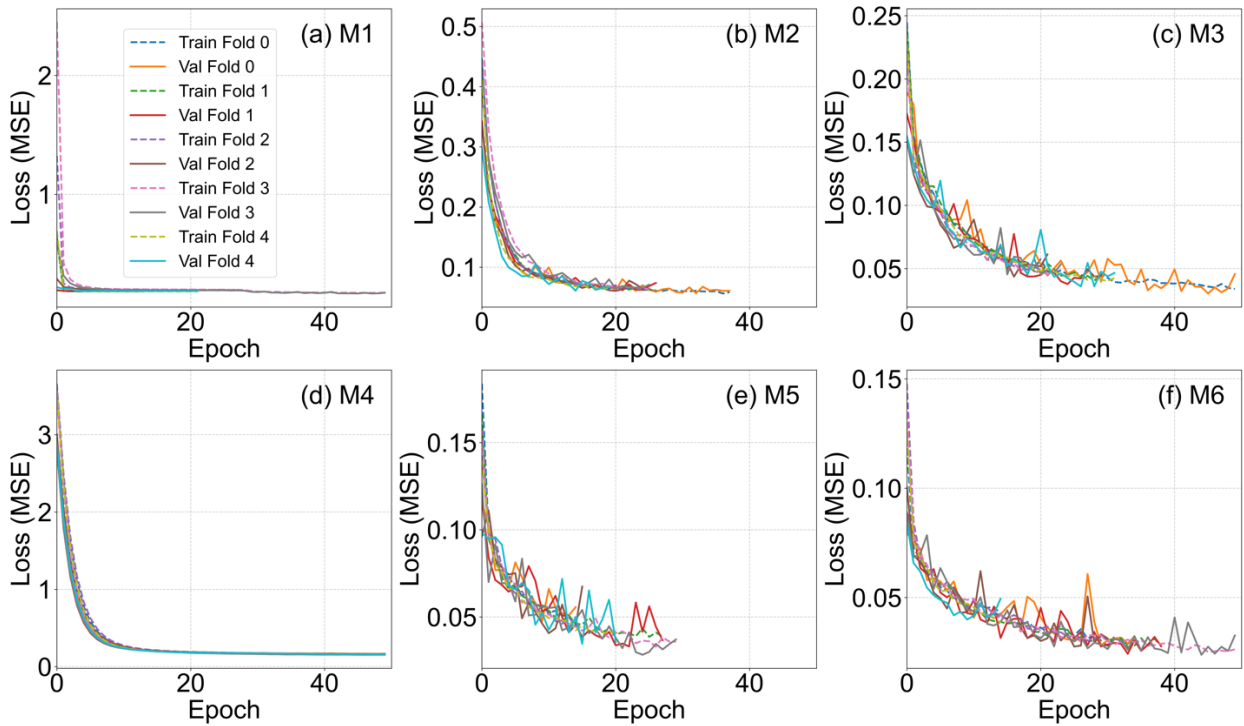

Fig. 12. Training and validation loss curves (MSE) across 5-fold cross-validation for the MLP models optimized using the scalarized loss function combining IoU and SSIM. While model selection was based on the scalarized validation score, MSE served as the internal loss function during training within each fold, using default settings of 50 epochs and a batch size of 256. All models exhibited convergence within 40–50 epochs. Higher variance in validation loss was observed for materials M3, M5, and M6, as shown in panels 11(c), (e), and (f), respectively.

Fig. 12 presents the training and validation loss curves (MSE) across 5-fold cross-validation for the scalarized loss function-based MLP models. All six material-specific models exhibit convergence within approximately 40–50 epochs, indicating that the optimization process was generally stable and effective despite the limited dataset size. However, materials such as M3, M5, and M6 display higher variance in validation loss across folds, suggesting greater sensitivity to fold composition and reduced generalization capacity. These fluctuations indicate performance disparities in the susceptibility predictions and reflect underlying challenges posed by input space sparsity and material domain shifts—issues that are explored in greater depth in the following section.

*3.3.4. Model Comparison and Generalization Limitations*

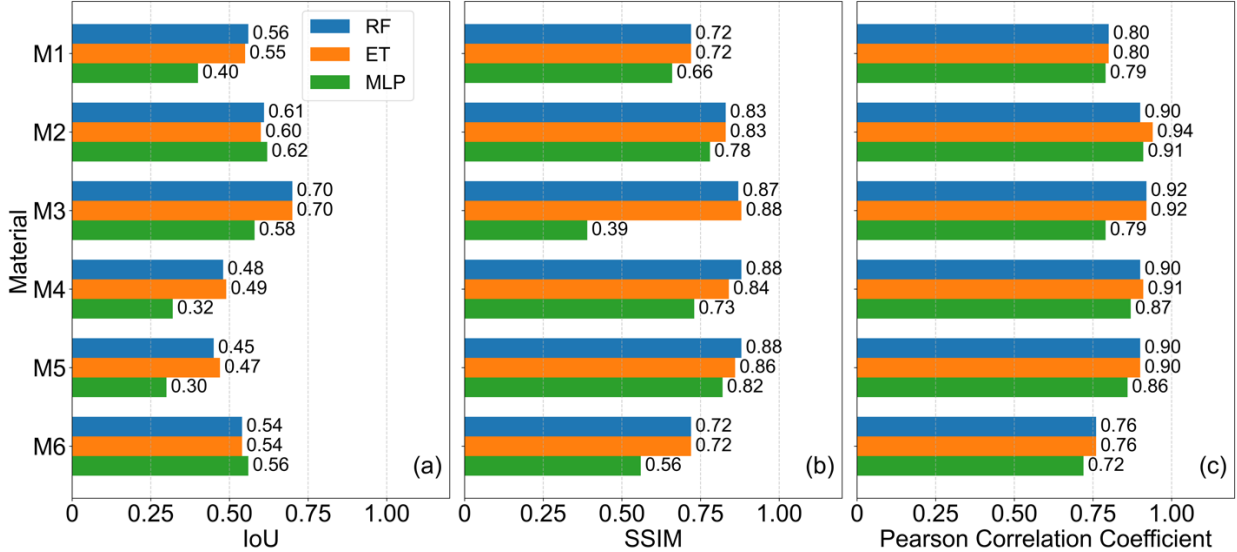

Fig. 13: Comparison of model performance across materials for RF (blue bars), ET (orange bars), and scalarized-loss MLP (green bars) models. Metrics shown are (a) IoU, (b) SSIM, and (c) Pearson correlation coefficient.

Fig. 13 presents a comparative performance analysis of the best-performing tree-based models, Random Forest (RF) and Extra Trees (ET) and the MLP model trained using a scalarized objective function across all six materials. While the MLP achieves competitive Pearson correlation scores, indicating it captures the overall trend in $\delta_{avg}$ reasonably well, it consistently underperforms both RF and ET in terms of IoU and SSIM. This performance gap can be explained by examining the Principal Component Analysis (PCA) plot in Fig. 14(a), which projects all input features of susceptibility data points (e.g., each $(V_{rf}, fd, \delta_{max0}, E_1, E_2)$ sample) into the first two principal components—linear combinations of the original features that capture the directions of maximum variance in the input space. These components were computed using the PCA class from the scikit-learn Python library with *n_components* = 2. The projection reveals that susceptibility data points from different materials form distinct, non-overlapping clusters in the reduced PCA space due to variations in material dependent SEY features. Under the leave-one-material-out cross-validation strategy, the held-out material occupies a region in feature space that is entirely disjoint from the training data. This presents a significant generalization challenge, particularly for neural networks like MLPs, which learn smooth, global function approximations and are known to struggle with out-of-distribution generalization when the training data lacks coverage of the test domain, as discussed by Arjovsky et al., 2020; Gulrajani and Lopez-Paz, 2020; Mahdavi et al., 2022. In contrast, ensemble tree models like Random Forests, due to their localized decision structures, often exhibit more stable performance in heterogeneous or sparsely sampled domains (Fernández-Delgado et al., 2014; Zhou and Feng, 2017).

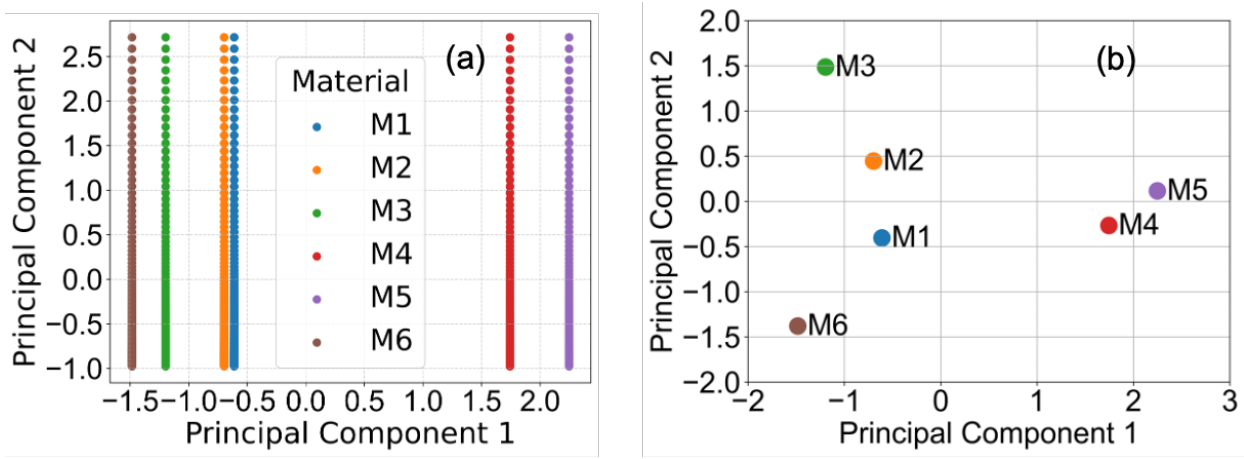

Fig. 14. Principal Component Analysis (PCA) of the input feature space. (a) Projection of all susceptibility data points defined by ($V_{rf}$, fd, $\delta_{max0}$, $E_1$, $E_2$) onto the first two principal components, color-coded by material, showing distinct, non-overlapping clusters. (b) PCA of the material dependent SEY parameters ($\delta_{max0}$, $E_1$, $E_2$) showing relative positioning of the six materials in the reduced feature space.

Building on this analysis, improving MLP performance should prioritize addressing the out-of-distribution (OOD) challenge posed by the sparse and disjoint material clusters in the feature space. The current hyperparameter optimization already spans deep architectures with up to 10 layers and 1024 neurons in the initial layer, making the MLP sufficiently expressive. Further increasing complexity risks overfitting to the limited training distribution and learning spurious correlations rather than generalizable physics. Instead, future efforts should prioritize expanding the dataset to include additional material property sets that populate the feature space more uniformly, thereby reducing the extrapolation burden and enabling neural networks to learn more robust, transferable patterns.

Nevertheless, our results confirm that the use of a scalarized objective—combining IoU and SSIM—is essential in enabling the neural network to learn physically meaningful susceptibility distributions, a feat not achieved with single-metric optimization. These results underscore that more sophisticated multi-objective optimization strategies could be a promising direction for future research.

Another important observation in our study is the consistent underperformance of all machine learning models when predicting susceptibility charts for materials M4 and M5, relative to the other materials. This can be explained from Fig. 14(b) which shows a PCA projection of the six material-dependent feature vectors, each defined by $\delta_{max0}$, $E_1$, and $E_2$. Materials M1, M2, M3, and M6 form a relatively compact cluster in the reduced space, indicating similar SEY characteristics. In contrast, M4 and M5 form a second cluster that is spatially offset and well-separated from the first, reflecting their pronounced divergence in material properties. This separation has direct implications for model generalization: under the leave-one-material-out strategy, when either M4 or M5 is excluded from training, the remaining data fail to adequately represent the held-out material. As a result, the models face a greater extrapolation burden and consistently exhibit degraded performance, as seen in the lower IoU and SSIM scores for M4 and M5 across all machine learning methods (Fig. 13(a)). These findings highlight a key limitation of this proof-of-concept study: the small and sparsely distributed dataset constrains model robustness. Expanding the dataset with a more diverse range of material property sets would be essential to support generalizable machine learning models for multipactor prediction.

## 4. Conclusion

This study establishes a supervised machine learning framework for predicting multipactor susceptibility in two-surface planar geometries, leveraging 3D electromagnetic Particle-in-Cell (PIC) simulation data generated for six materials with distinct secondary electron yield (SEY) characteristics. We systematically evaluated linear regression models, nonlinear tree-based ensemble methods, and neural network architectures using a leave-one-material-out validation scheme to assess cross-material generalization. Our findings reveal that:

1) Linear regression models are inadequate for predicting multipactor susceptibility due to their inability to capture the strong nonlinear relationships between input features and electron growth behavior.
2) Tree-based ensemble methods (RF and ET) consistently outperform other models in preserving the spatial and structural features of susceptibility charts, owing to their localized partitioning and robust generalization across disjoint material domains.
3) Multilayer Perceptrons (MLPs), even when optimized with scalarized loss functions, trail behind tree-based models, primarily due to their sensitivity to out-of-distribution data under sparse training conditions.
4) Scalarized loss functions combining IoU and SSIM significantly improve neural network performance compared to solo-objective optimization, suggesting that multi-objective learning is a promising path forward.
5) Principal Component Analysis (PCA) reveals that poor generalization—especially for materials M4 and M5—stems from feature-space isolation, highlighting the need for more comprehensive material coverage.

In sum, this work not only demonstrates the feasibility of applying machine learning to model multipactor susceptibility but also identifies critical directions for advancing this approach: (1) expanding the material dataset to alleviate extrapolation challenges and improve generalization; (2) integrating uncertainty quantification to assess predictive reliability; and (3) developing physics-informed or hybrid models to enhance robustness and interpretability. Pursuing these avenues will be vital for transitioning ML-based multipactor prediction from a proof of concept into practical tools for the design and diagnostics of high-power RF and accelerator systems.

Declaration of Generative AI and AI-assisted technologies in the writing process: During the preparation of this work ChatGPT (GPT-4-turbo, developed by OpenAI) was used in order to assist with grammatical refinement and refinement of scientific language. After using this tool, the authors reviewed and edited the content as needed and take full responsibility for the content of the publication.

This work was supported by the Air Force Office of Scientific Research (AFOSR) MURI under Grant No. FA9550-21-1-0367, the Air Force Office of Scientific Research (AFOSR) MURI Grant No. FA955020-1-0409, and the Air Force Office of Scientific Research (AFOSR) under Grant No. FA9550-22-1-0523.